\documentclass{aa}             

\input epsf

\begin{document}

\def\24{V1130\,Tau}
\def\kms{\ifmmode{\rm km\thinspace s^{-1}}\else km\thinspace s$^{-1}$\fi}
\def\cm{\ifmmode{\rm cm\thinspace^{-1}}\else cm\thinspace$^{-1}$\fi}
\def\Msun{$M_{\sun}$}
\def\Rsun{R_{sun}}
\def\Lsun{L_{sun}}
\def\feh{$[\mathrm{Fe/H}]$}
\def\meh{$[\mathrm{M/H}]$}
\def\afe{$[\mathrm{\alpha/Fe}]$}
\def\ione{\,{\sc i}}
\def\itwo{\,{\sc ii}}

\title{
Absolute dimensions of eclipsing binaries. 
\thanks{Based on observations carried out at the Str{\"o}mgren Automatic
Telescope (SAT) and the 1.5m telescope at ESO, La Silla
(62.H-0319, 62.L-0284, 63.H-0080, 64.L-0031, 66.D-0178)
}}
\subtitle{XXVII. V1130\,Tauri: \\
A metal-weak F-type system, perhaps with preference for $Y = 0.23-0.24$
\thanks{
Table~11 is available in electronic form
at the CDS via anonymous ftp to cdsarc.u-strasbg.fr (130.79.128.5)
or via http://cdsweb.u-strasbg.fr/cgi-bin/qcat?J/A+A/}
}
\author{
J.V. Clausen     \inst{1}
\and
E.H. Olsen       \inst{1}
\and
B.E. Helt        \inst{1}
\and
A. Claret        \inst{2}
}
\offprints{J.V.~Clausen, \\ e-mail: jvc@nbi.ku.dk}

\institute{
Niels Bohr Institute, Copenhagen University,
Juliane Maries Vej 30,
DK-2100 Copenhagen {\O}, Denmark
\and
Instituto de Astrof\'isica de Andaluc\'ia, CSIC,
Apartado 3004, E-18080 Granada, Spain
}

\date{Received 20 November 2009 / Accepted 4 December 2009} 

\titlerunning{V1130\,Tau}
\authorrunning{J.V. Clausen et al.}

\abstract
{Double-lined, detached eclipsing binaries are our main source for accurate
stellar masses and radii.
This paper is the first in a series with focus on the upper half of the 
main-sequence band and tests of 1--2 \Msun\ evolutionary models. 
}
{We aim to determine absolute dimensions and abundances for the
detached eclipsing binary \24, and to perform a detailed
comparison with results from recent stellar evo\-lu\-tio\-nary models.}
{$uvby$ light curves and $uvby\beta$ standard photometry have been obtained with
the Str\"omgren Automatic Telescope, and high-resolution spectra
have been acquired at the FEROS spectrograph; both are ESO, La Silla facilities.
We have applied the Wilson-Devinney model for the photometric analysis,
spectroscopic elements are based on radial velocities measured via 
broadening functions, and \feh\ abundances have been determined from synthetic 
spectra and $uvby$ calibrations.
}
{\24 is a bright ($m_{V}$ = 6.56), nearby ($71 \pm 2$ pc) detached system with 
a circular orbit ($P = 0\fd80$). The components are deformed with filling factors 
above 0.9. Their masses and radii have been established to 0.6--0.7\%.
We derive a \feh\ abundance of $-0.25 \pm 0.10$.
The measured rotational velocities, $92.4 \pm 1.1$ (primary) and
$104.7 \pm 2.7$ (secondary) \kms, are in fair agreement with synchronization.
The larger 1.39 \Msun\ secondary component has evolved to the middle of the 
main-sequence band and is slightly cooler than the 1.31 \Msun\ primary.
Yonsai-Yale, BaSTI, and Granada evolutionary models for the observed
metal abundance and a 'normal' He content of $Y = 0.25-0.26$, 
marginally reproduce the components at ages between 1.8 and 2.1 Gyr.
All such models are, however, systematically about 200 K hotter than observed 
and predict ages for the more massive component, which are systematically
higher than for the less massive component.
These trends can not be removed by adjusting the amount of core overshoot or 
envelope convection level, or by including rotation in the model calculations.
They may be due to proximity effects in \24, but on the other hand, we find 
excellent agreement for 2.5--2.8 Gyr Granada models with a slightly lower 
$Y$ of 0.23--0.24. 
}
{\24\ is a valuable addition to the very few well-studied 1--2 \Msun\ binaries
with component(s) in the upper half of the main-sequence band, or beyond.
The stars are not evolved enough to provide new information on the dependence
of core overshoot on mass (and abundance), but might - together with a larger sample 
of well-detached systems - be useful for further tuning of the helium enrichment law.
Analyses of such systems are in progress.
}

\maketitle

\section{Introduction}
\label{sec:intro}

In this paper, we present the first detailed study of the
bright ($m_{V}$ = 6.56), early F-type, double-lined eclipsing 
binary \object{V1130\,Tau}.
The orbital period is short ($P = 0\fd80$), but the system is
still detached, and for several reasons it is an interesting case.
First, it is more evolved than most of the well studied early F-type 
main sequence systems; actually the more massive, larger component 
has become the slightly cooler one. Next, it is reported to be metal-weak. 
Finally, it is situated at a distance of only 71 pc, meaning
that it belongs to the (small) group of eclipsing binaries within 125 pc,
discussed by Popper (\cite{dmp98}), which could be useful for improving the
radiative flux scale.

In the following, we determine absolute dimensions and
abundances, based on analyses of new $uvby$ light curves and high-resolution 
spectra, and compare \24\ to several modern stellar evolutionary models.
Throughout the paper, the component eclipsed at the slightly deeper eclipse 
at phase 0.0 is referred to as the primary $(p)$, and the other as the 
secondary $(s)$ component.

\begin{table*}
\caption[]{\label{tab:24133_std}
Photometric data for \24\ and the comparison stars.}
\begin{minipage}{\textwidth}
\begin{center}
\begin{tabular}{lllrrrrrrrrrrrr} \hline
\hline\noalign{\smallskip}
Object&Sp. Type&Ref.&$V$&$\sigma$&$b-y$&$\sigma$&$m_1$&$\sigma$&$c_1$ &$\sigma$&N($uvby$)&$\beta$&$\sigma$&N($\beta$)\\
\noalign{\smallskip}
\hline
\noalign{\smallskip}
V1130\,Tau& F2\,V$^{\mathrm{a}}$  & C10 & 6.556 & 9& 0.263 & 3& 0.140 & 7& 0.478 & 9& 19 & 2.653 & 6 &32\\   
          &        & F88 & 6.594 &30& 0.272 & 0& 0.136 & 2& 0.478 & 1&  2 & 2.652 &   & 1\\
          &        & O83 & 6.639 &50& 0.276 & 1& 0.124 & 4& 0.474 & 3&  4 & 2.652 & 7 & 3\\
\hline
 HD23503  & F2/3\,V$^{\mathrm{a}}$& C10 & 8.262 & 6& 0.261 & 5& 0.174 & 8& 0.509 & 9&156 & 2.686 & 8 &22\\ 
          &        & O83 & 8.251 & 1& 0.272 & 0& 0.163 & 9& 0.510 & 1&  2 &       &   &  \\
          &        & O94 & 8.264 & 5& 0.266 & 3& 0.167 & 4& 0.497 & 6&  1 & 2.666 & 6 & 1\\
\hline
 HD24552  & G1\,V$^{\mathrm{a}}$  & C10 & 7.979 & 6& 0.386 & 5& 0.200 & 8& 0.325 & 8&107 & 2.596 & 8 &16\\
          &        & O83 & 7.972 & 6& 0.392 & 3& 0.200 & 1& 0.329 & 3&  2 &       &   &  \\
          &        & O94 & 7.975 & 5& 0.387 & 3& 0.196 & 4& 0.326 & 6&  1 & 2.575 & 6 & 1\\
\hline
 HD25059  & G3\,V$^{\mathrm{a}}$  & C10 & 9.161 & 5& 0.391 & 4& 0.192 & 8& 0.329 & 9&101 & 2.591 & 9 &19\\
          &        & O93 &       &  & 0.399 & 3& 0.164 & 5& 0.359 & 6&  1 &       &   &  \\
          &        & O94 & 9.163 & 4& 0.392 & 3& 0.185 & 5& 0.335 & 7&  1 & 2.583 & 6 & 1\\
\hline
\end{tabular}
\begin{list}{}{}
\item[$^{\mathrm{a}}$] Houk \& Swift (\cite{houk99}); for \24, see also spectral types in Sect.~\ref{sec:24133}
\end{list}
\end{center}
\textsc{NOTE 1:}
References are:
C10 = This paper,
F88  = Franco (\cite{franco88}), 
O83  = Olsen (\cite{olsen83}),   
O93  = Olsen (\cite{olsen93}),   
O94  = $uvby$: Olsen (\cite{olsen94}), $\beta$: Olsen (unpublished).   

\textsc{NOTE 2:}
For \24, the $uvby$ information by C09 is the mean value at phases 0.25 and 0.75,
and the $\beta$ information is the mean value outside eclipses.

\textsc{NOTE 3:}
N is the total number of observations used to form the mean values, and
$\sigma$ is the rms error (per observation) in mmag.
\end{minipage}
\end{table*}

\section{V1130\,Tau}
\label{sec:24133}

HD~24133 (CSV~356, HIP~17988) was confirmed to be variable by Olsen
(\cite{olsen83}). Based on the $uvby$ photometry, Olsen supplied Abt with a list
containing about 800 potentially weak-lined A5-G0 stars, and Abt (\cite{abt86})
subsequently classified HD~24133 as F3\,V\,wl (A5 met) and also reported it
to be a double-lined spectroscopic binary.
Gray (\cite{gray89}) classified HD~24133 as F5\,V\,m-2, Gray \& Garrison
{\cite{graygar89}) confirmed that its metal lines have the strength
of an A5 star, and Gray et al. (\cite{gray01}) found it to be
a fairly rapid rotator, but clearly metal-weak. 
The eclipsing nature 
was discovered by Hipparcos (ESA \cite{hip97}; orbital period
$0\fd7988710$), and HD~24133 was subsequently
assigned the variable name \24\ (Kazarovets et al. \cite{k99}).
Rucinski et al. (\cite{r03}) determined a spectroscopic orbit,
leading to $(M_p + M_s)\mathrm{sin^3}i = 2.41 \pm 0.03$ \Msun, 
and noted that \24\ is one of the shortest period detached early 
F-type systems.
Besides a few times of minima, nothing has been published on this
binary since then.

\section{Photometry}
\label{sec:phot}

Below, we present the new photometric material for \24\ and refer to Clausen
et al. (\cite{jvcetal01}; hereafter CHO01) for further details on observation
and reduction procedures, and determination of times of minima.

\subsection{Light curves for V1130\,Tau}
\label{sec:lc}

\begin{figure*}
\epsfxsize=185mm
\epsfbox{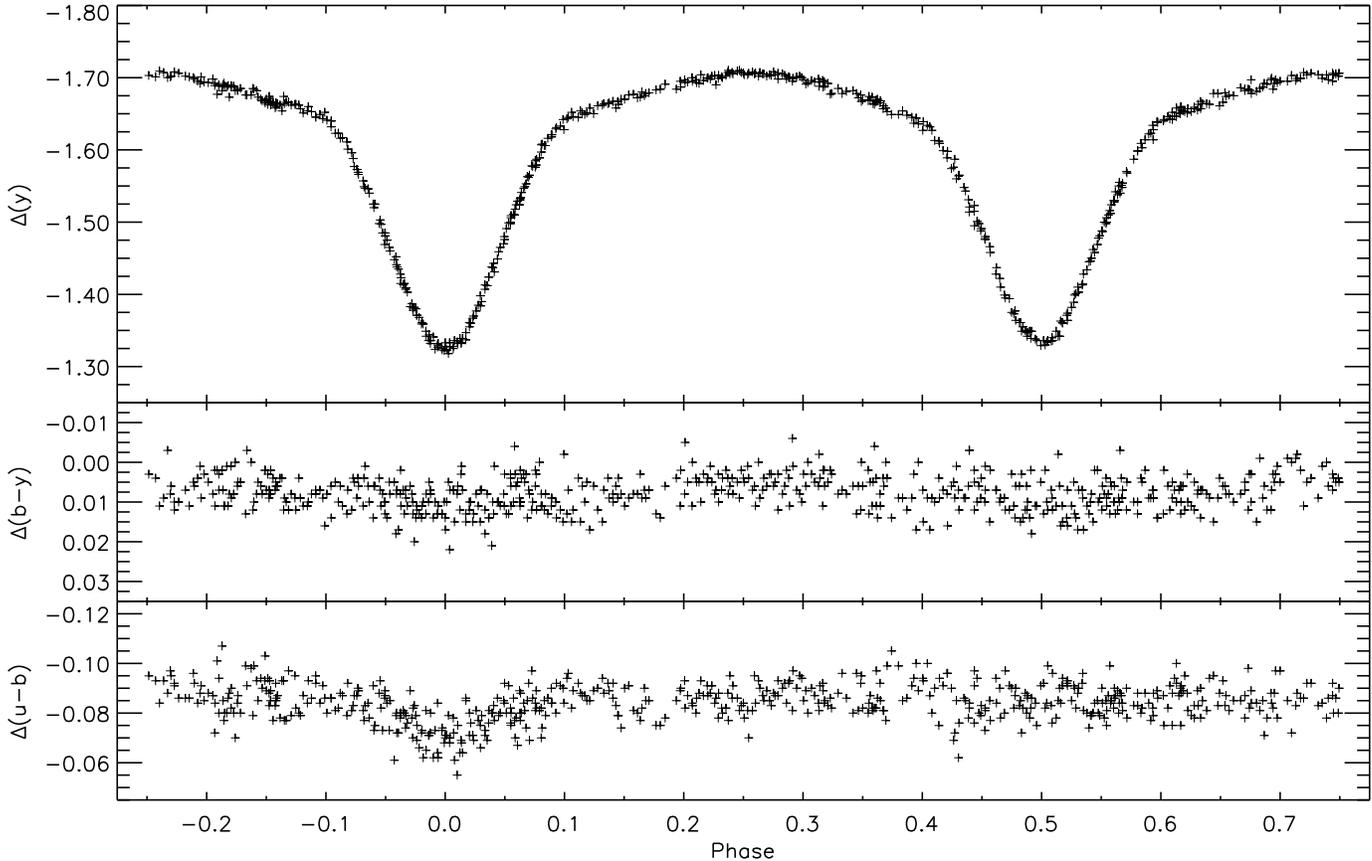}
\caption[]{\label{fig:24133_lc}
$y$ light curve and $b-y$ and $u-b$ colour curves (instrumental system)
for V1130\,Tau.
}
\end{figure*}

The differential $uvby$ light curves of \24\ were observed at the
Str{\"o}mgren Automatic Telescope (SAT) at ESO, La Silla
and its 6-channel $uvby\beta$ photometer on 59 nights between
October 1997 and November 1998 (JD2450727--2451120).
They contain 583 points per band with all phases covered at least twice.
The observations were done through an 18 arcsec diameter circular
diaphragm at airmasses between 1.2 and 1.8.
\object{HD~23503}, \object{HD~24552}, and \object{HD~25059} -
all within a few degrees from \24\ on the sky - were used as
comparison stars and were all found to be constant within a few mmag;
see Table~\ref{tab:24133_std}.
The light curves were calculated relative to HD~23503, but all comparison
star observations were used, shifting them first to the same light level.
The average accuracy per point is about 4--5 mmag ($ybv$) and 7 mmag ($u$).

As seen from Fig.~\ref{fig:24133_lc}, \24\ is detached but fairly
close, with $y$ eclipse depths of about 0.4 mag. Primary eclipse is
only marginally deeper than secondary, meaning that the surface fluxes
of the components are nearly identical.
The light curves (Table 11) will only be available in electronic form.

\subsection{Standard photometry for \24}
\label{sec:std}

Standard $uvby\beta$ indices for \24\ (between eclipses) and the 
three comparison stars,
observed and derived as described by CHO01, are presented in
Table~\ref{tab:24133_std}. The indices are based on many
observations and their precision is high.
For comparison, we have included published photometry from
other sources. In general, the agreement is good, 
but individual differences larger than the quoted errors occur.

\subsection{Times of minima and ephemeris for \24}
\label{sec:tmin}

Three times of each of primary and secondary minimum have been established
from the $uvby$ light curve observations; see Table~\ref{tab:24133_tmin}.
A list of earlier times of minima was kindly provided by
Kreiner; see Kreiner et al. (\cite{kreiner01}) and
Kreiner (\cite{kreiner04})\footnote{{\scriptsize\tt http://www.as.ap.krakow.pl/ephem}}.
Except for two unpublished times based on Hipparcos photometry,
which showed large deviations, they were included in the ephemeris analysis 
together with the recently published time of primary minimum by
Brat et al. (\cite{brat08}). 

Assuming a circular orbit, we derive the following linear ephemeris
from a weighted least squares fit to all accepted times of minima:

\begin{equation}
\label{eq:24133_eph}
\begin{tabular}{r r c r r}
{\rm Min \, I} =  & 2450770.69601 & + & $0\fd 798868143$ &$\times\; E$ \\
                  &      $\pm   9$&   &    $\pm      38$ &             \\
\end{tabular}
\end{equation}

\noindent
Separate weighted linear least squares fits to the times of 
primary and secondary minima lead to identical orbital periods, and we
adopt Eq.~\ref{eq:24133_eph} for the analyses of the $uvby$ light curves 
and radial velocities.

\begin{table}
\caption[]{\label{tab:24133_tmin}Times of primary (P) 
and secondary (S) minima of V1130\,Tau
determined from the $uvby$ observations.
}
\begin{minipage}{\columnwidth}
\centering
\renewcommand{\footnoterule}{}  
\begin{tabular}{cccr} \hline
\hline\noalign{\smallskip}
HJD           & rms    & Type &   O-C\footnote{Calculated from the ephemeris given in Eq.~\ref{eq:24133_eph}}\\
$-$ 2\,400\,000 &      &      &            \\
\noalign{\smallskip}
\hline
\noalign{\smallskip}
50742.73603   &0.00040 & P    &   0.00041  \\
50770.69594   &0.00020 & P    & $-0.00007$ \\
50778.68480   &0.00020 & P    &   0.00011  \\
50774.73329   &0.00040 & S    &   0.00049  \\
50776.68754   &0.00020 & S    &   0.00002  \\
50780.68191   &0.00020 & S    &   0.00005  \\
\hline
\end{tabular}
\end{minipage}
\end{table}

\subsection{Photometric elements}
\label{sec:phel}

\begin{table}
\caption[]{\label{tab:24133_wd}
Photometric solutions for \24.
}
\begin{center}
\begin{tabular}{lrrrr} \hline
\hline\noalign{\smallskip}
                     &     $y$    &       $b$  &       $v$  &   $u$\\                   
\noalign{\smallskip}
\hline
\noalign{\smallskip}
$i$ \, (\degr)       &  73.97     &   73.84    &   73.73    &  74.20\vspace{-0.8mm}\\   
                     & $\pm 2$    &  $\pm 4$   &  $\pm 4$   & $\pm 6$\\                 

$\Omega_p$           &   4.5138   &    4.5079  &    4.4954  &   4.5642\vspace{-0.8mm}\\  
                     &  $\pm 39$  &   $\pm 40$ &   $\pm 41$ &  $\pm 64$\\                

$\Omega_s$           &   4.0727   &    4.0753  &    4.0775  &   4.0805\vspace{-0.8mm}\\  
                     &  $\pm 25$  &   $\pm 29$ &   $\pm 30$ &  $\pm 45$\\                

$r_p$                &  0.2956    &   0.2961   &   0.2973   &  0.2911\\                 
                                                            
$r_s$                &  0.3542    &   0.3539   &   0.3536   &  0.3532\\                 
                                                                                        
$k$                  &  1.198     &   1.195    &   1.189    &  1.213\\                  
                                                                                        
$r_p + r_s$          &  0.6498    &   0.6500   &   0.6509   &  0.6443\\                 

$u_p = u_s$          &  0.52      &   0.61     &   0.67     &  0.63   \\
                                                                                        
$T_{{\rm eff},s}$    &  6638      &   6639     &   6643     &  6628\vspace{-0.8mm}\\
                     &$\pm 4$     & $\pm 3$    &$\pm  3$    &$\pm 4$\\

$L_s/L_p$            &  1.444     &    1.436   &   1.429    &  1.436  \\

$\sigma$ \, (mmag.)  &  4.5       &    4.3     &   4.3      &  6.5   \\
\noalign{\smallskip}            
\hline
\end{tabular}            
\end{center}            
\textsc{Note 1:}
Limb darkening coefficients by van Hamme (\cite{vh93}),
gravity darkening exponents of 0.33, and bolometric albedo
coefficients of 0.5 were adopted, as appropriate for convective envelopes.
$T_{{\rm eff},p}$ was assumed to be 6650 K, see Sect.~\ref{sec:absdim}.

\textsc{Note 2:}
The errors quoted for the free parameters are the $formal$ standard errors
determined from the iterative least squares solution procedure.
\end{table}                       

\begin{table}            
\caption[]{\label{tab:24133_phel}
Adopted photometric elements for \24.
}
\begin{center}             
\begin{tabular}{ll}             
\noalign{\smallskip}             
\hline             
\noalign{\smallskip}             
$i$              & $73{\fdg}82 \pm 0{\fdg}20$\\
$r_p$            & $0.2952 \pm 0.0018$\\      
$r_p(pole)$      &  0.2862\\
$r_p(point)$     &  0.3125\\
$r_p(side)$      &  0.2935\\
$r_p(back)$      &  0.3048\\
$r_s$            & $0.3535 \pm 0.0021$\\ 
$r_s(pole)$      &  0.3369\\
$r_s(point)$     &  0.3965\\
$r_s(side)$      &  0.3506\\
$r_s(back)$      &  0.3717\\
\noalign{\smallskip}             
\end{tabular}             
\begin{tabular}{lrrrr}             
\noalign{\smallskip}             
          & $y$    & $b$    & $v$   & $u$\\           
\noalign{\smallskip}
$L_s/L_p$ &  1.443  &  1.437  &   1.432  &   1.434\vspace{-0.8mm}\\
          &$\pm  23$&$\pm  23$&$\pm   23$ &$\pm  23$\\ 
\noalign{\smallskip}             
\hline             
\end{tabular}             
\end{center}
\textsc{Note:}
The individual luminosity ratios are based on the mean stellar and 
orbital parameters.
\end{table}

Since the relative radii of the components of \24\ are fairly large,
0.25--0.30, we have adopted the Wilson-Devinney model (Wilson \& Devinney 
\cite{rew1971}; Wilson \cite{rew1979, rew1990, rew1993}; Van Hamme \& Wilson 
\cite{rew2003}) for the light curve analyses. We have used the JKTWD code
developed by J. 
Southworth\footnote{{\scriptsize\tt http://www.astro.keele.ac.uk/$\sim$jkt/}},
which is based on the 2003 version of the 'Binary Star Observables Program'
by Wilson et al.\footnote{{\scriptsize\tt ftp://ftp.astro.ufl.edu/pub/wilson/}} 
This code was recently applied for the light curve
analyses of \object{DW\,Car} (Southworth \& Clausen \cite{sc07})
and \object{V380\,Cyg} (Pavlovski et al. \cite{p09}).

Mode 2 (detached binaries) was used throughout, and the
stellar atmosphere approximation functions for the $uvby$ bands
were adopted (Van Hamme \& Wilson \cite{rew2003},
Kurucz \cite{ku93}). 
The effective temperature of the primary component was kept
at 6500~K; see Sect.~\ref{sec:absdim}.
A linear limb darkening law was  assumed with coefficients
adopted from Van Hamme (\cite{vh93}). 
The linear coefficients by Claret (\cite{c00}) are about 0.1 larger
and lead to a $0\fdg2$ lower orbital inclination, whereas the 
radii are practically unchanged. Within errors, non-linear limb darkening 
lead to identical photometric elements. 
Gravity darkening exponents corresponding to convective
atmospheres were applied, and bolometric reflection albedo coefficients 
of 0.5 were chosen, again due to convection. The simple reflection mode
(MREF = 1) was used; we note that the detailed mode gives nearly identical elements.
The mass ratio between the components was kept at the spectroscopic value
($M_s/M_p=1.066\pm0.004$), and synchronous rotation was assumed.
The light curves were analysed independently with at least 10 differential 
parameter corrections, and continuing until they were below 20\% of the 
corresponding formal standard errors for all parameters. 
The stability of the adopted solutions was tested by adding 100 more iterations.

In tables and text, we use the following symbols:
$i$ orbital inclination;
$r$ relative volume radius;
$k = r_s/r_p$;
$\Omega$ surface potential;
$u$ linear limb darkening coefficient;
$L$ luminosity;
$T_{\rm eff}$ effective temperature.

The individual solutions are presented in Table~\ref{tab:24133_wd},
and $O\!-\!C$ residuals of the $b$ observations from the theoretical 
light curve are shown in Fig.~\ref{fig:24133_res_b}. Equally good
fits are obtained in the three other bands, and the rms of the residuals
correspond closely to the observational accuracy.
As expected, the less massive, slightly hotter primary component is
also the smaller one. 
\24\ is detached, but the filling factors of the components are above 0.9,
and the deformation of the secondary component is significant.
The relative volume radii obtained from the four
bands agree well, except perhaps for the less precise $u$ result 
for the primary. 
We find no evidence of third light, neither in the spectra nor from
the light curve solutions, and the small differences in orbital inclinations 
are probably due to model and/or limb darkening effects.
The adopted photometric elements are listed in Table~\ref{tab:24133_phel}
with realistic uncertainties, which reflect the formal standard errors and
the interagreement of the $uvby$ results, and also take into account the 
consequences of $\pm 0.1$ changes of limb darkening coefficients and the 
uncertainly of the mass ratio. 
As seen, the relative volume radii have been established to about 0.6\%. 

\begin{figure}
\epsfxsize=85mm
\epsfbox{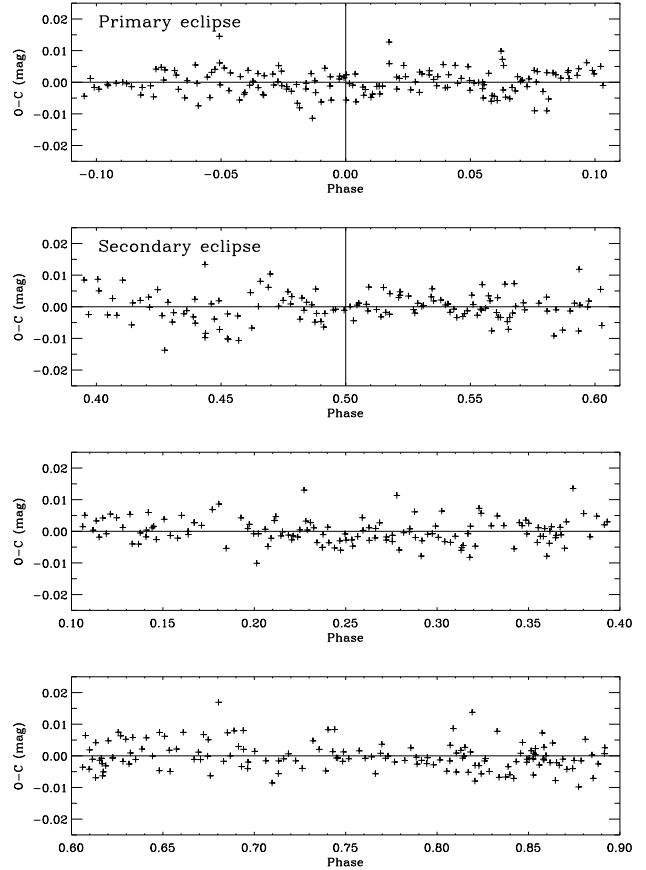}
\caption[]{\label{fig:24133_res_b}
($O\!-\!C$) residuals of the \24\ $b$-band observations from the theoretical
light curve computed for the photometric elements given in
Table~\ref{tab:24133_wd}.
}
\end{figure}

\section{Spectroscopy}
\label{sec:spec}

\subsection{Spectroscopic observations}
\label{sec:specobs}

For radial velocity and abundance determinations, we have obtained 
18 high-resolution spectra with the FEROS fiber 
echelle spectrograph at the ESO 1.52-m telescope at La Silla, Chile 
(Kaufer et al. \cite{feros99}, \cite{feros00}).
The spectrograph, which resides in a temperature-controlled room,
covers without interruption the spectral region from the Balmer jump
to 8700\,\AA, at a constant velocity resolution of 2.7 \kms\
per pixel ($\lambda/\Delta\lambda=48\,000$).
We refer to Clausen et al. (\cite{avw08}, hereafter CTB08) 
for details on the reduction of the spectra, which
were observed between January 1999 and March 2001; 
an observing log is given in Table~\ref{tab:feros}.

\begin{table}
\caption[]{\label{tab:feros}
Log of the FEROS observations of \24.
}
\begin{minipage}{\columnwidth}
\centering
\renewcommand{\footnoterule}{}  
\begin{tabular}{ccrr}
\hline
\hline\noalign{\smallskip}
HJD$-$2\,400\,000\footnote{Refers to mid-exposure}  & phase &t$_{exp}$\footnote{Exposure time in seconds} & 
S/N\footnote{Signal-to-noise ratio measured around 6070 {\AA}. At the shorter wavelengths used for the radial velocity measurements it is somewhat lower}\\
\noalign{\smallskip}
\hline
\noalign{\smallskip}
 51188.6379 &  0.1675 &    600 &   223 \\ 
 51207.5446 &  0.8344 &    600 &   171 \\ 
 51207.5662 &  0.8615 &    600 &   170 \\ 
 51208.5852 &  0.1370 &    600 &   162 \\ 
 51209.5918 &  0.3971 &    600 &   200 \\ 
 51211.5630 &  0.8645 &    600 &    84 \\ 
 51212.6064 &  0.1706 &    600 &   176 \\ 
 51385.9274 &  0.1288 &    600 &   142 \\ 
 51386.9327 &  0.3872 &    600 &   170 \\ 
 51390.8696 &  0.3154 &    600 &   241 \\ 
 51391.9171 &  0.6266 &    600 &   225 \\ 
 51392.8746 &  0.8252 &    720 &   237 \\ 
 51562.5335 &  0.1992 &    600 &   243 \\ 
 51562.5962 &  0.2777 &    600 &   281 \\ 
 51562.6289 &  0.3187 &    600 &   244 \\ 
 51977.4958 &  0.6371 &    630 &   181 \\ 
 51978.4897 &  0.8812 &    600 &   179 \\ 
 51981.4927 &  0.6402 &    600 &   166 \\ 
\hline
\end{tabular}
\end{minipage}
\end{table}

\subsection{Radial velocities}
\label{sec:rv}

\begin{table}
\caption[]{\label{tab:24133_orders}
Echelle orders and wavelength ranges used for the radial 
velocity measurements of \24.
}
\begin{center}
\begin{tabular}{crcr}
\hline
\hline\noalign{\smallskip}
Order  & Range ({\AA}) &Order  & Range ({\AA}) \\
\noalign{\smallskip}
\hline
\noalign{\smallskip}
 55 & 4020 - 4090 &   53 & 4170 - 4240 \\ 
 52 & 4220 - 4330 &   50 & 4395 - 4500 \\ 
 49 & 4510 - 4590 &   45 & 4905 - 4975 \\ 
 44 & 4975 - 5090 &   43 & 5090 - 5212 \\ 
\hline
\end{tabular}
\end{center}
\end{table}

The radial velocities for \24\ were measured from eight useful orders
(4020 -- 5210 {\AA}) of the 18 FEROS spectra, see Table~\ref{tab:24133_orders}.
The selection of this limited number of orders was based on initial analyses, 
which showed that several orders give unreliable results and have to be excluded, 
either because too few lines are available, or because they contain defects or
are difficult to normalise properly.
We applied the broadening function (BF) formalism (Rucinski \cite{r99,r02,r04}),
using synthetic templates with no rotational broadening, calculated for
$T_{\rm eff} = 6650$ K, log($g$) = 4.0, and \feh\,$=-0.25$.
They were produced with the $bssynth$ tool (Bruntt, private
communication), which applies the SYNTH software (Valenti \& Piskunov
\cite{vp96}) and modified ATLAS9 models (Heiter \cite{heiter02}).
Line information was taken from the Vienna Atomic Line Database (VALD;
Kupka et al. \cite{kupka99}).
BF's were then produced for each of the selected orders of each spectrum.

\begin{figure}
\epsfxsize=80mm
\epsfbox{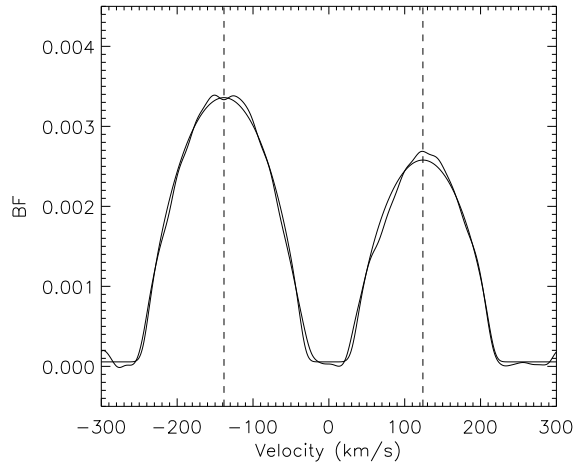}
\caption[]
{
\label{fig:24133_bf}
Broadening function (thick) obtained for the 5090-5212\,\AA\ region
and theoretical fit (thin).
The FEROS spectrum was taken at phase 0.834, at HJD$=$2451207.5446. 
The primary component is to the right.
}
\end{figure}

Due to the deformation of the components (Table~\ref{tab:24133_phel}) and
to reflection effects, the lineprofiles and thereby the
BF's can not a priori be expected to be symmetric.
The optimum method for radial velocity determinations is therefore
to fit theoretical BF's which take the proximity effects into account;
for W~UMa systems, see e.g. Rucinski et al (\cite{rls93}) and references 
therein. 
Such BF's were constructed, for both components at each phase and order, 
from synthetic rotational broadened line profiles calculated via the 
Wilson-Devinney (WD) model (Sect.~\ref{sec:phel}). The radial velocities
were then determined by fitting combined theoretical BF's to the observed
ones, i.e. by shifting and scaling the theoretical BF's for each component
until the best fit was obtained for the combination.

Since the observed BF's do not show clear asymmetries, we have also
applied an approach based on simpler theoretical BF's, which assume that the 
stars are spherical, rigid rotators (e.g. Kaluzny et al. \cite{k06}).
Before the radial velocity determination, these BF's were 
convolved with a Gauss profile corresponding to the instrumental resolution. 
In general, very good fits were obtained from this approach; 
see Fig.~\ref{fig:24133_bf}.

In the case of \24, the two BF approaches result in radial velocities 
for all spectral orders and phases, which
agree within about 1--2 \kms, and the differences do not correlate with 
orbital phase and/or velocity separation. In Sect.~\ref{sec:spel} we present
orbital solutions from both sets of radial velocities.

As described by e.g. Kaluzny et al. (\cite{k06}), the projected rotational
velocities $v \sin i$ of the components and (monochromatic)
light/luminosity ratios between them can also be obtained from analyses 
of the simple broadening functions mentioned above. We have tested this on 
synthetic binary spectra with input rotational velocities of 90.0 (primary) 
and 110.0 (secondary) \kms, corresponding closely to pseudosynchronous rotation,
and a light ratio of 1.44, and we find that the method is safe for \24.
The rotational velocities determined from the BF analyses are within 1 \kms 
from the input values, and the light ratio is reproduced to high precision. 
Analyses of the observed \24\ spectra yield mean $v \sin i$ velocities of
$92.4 \pm 1.1$ (primary) and $104.7 \pm 2.7$ \kms (secondary), and
the mean light ratio, $1.44 \pm 0.02$, is in perfect agreement with
the results from the light curve analyses (Table~\ref{tab:24133_phel}). 
No significant wavelength/order dependencies are seen.

\subsection{Spectroscopic elements}
\label{sec:spel}

\begin{table*}   
\caption[]{\label{tab:24133_spel}
Spectroscopic orbital solution for \24.
}
\begin{minipage}{\textwidth}
\centering
\renewcommand{\footnoterule}{}  
\begin{tabular}{lrrr} \hline   
\hline\noalign{\smallskip}    
Parameter&\multicolumn{1}{c}{WD based BF}&\multicolumn{1}{c}{Symmetrical BF}&\multicolumn{1}{c}{Mean velocities}\\ 
         &                              &                              &\multicolumn{1}{c}{Adopted}\\
\noalign{\smallskip}
\hline
\noalign{\smallskip}    
Adjusted quantities:            &   \\ 
$K_p$~(\kms)                    &$158.14 \pm 0.34$ & $158.33 \pm 0.37$ & $158.29 \pm 0.34$ \\
$K_s$~(\kms)                    &$148.45 \pm 0.42$ & $148.49 \pm 0.27$ & $148.52 \pm 0.33$ \\
$\gamma_p$~(\kms)               &$-10.98 \pm 0.27$ & $-11.12 \pm 0.30$ & $-11.05 \pm 0.28$ \\
$\gamma_s$~(\kms)               &$-11.03 \pm 0.34$ & $-11.41 \pm 0.22$ & $-11.22 \pm 0.27$ \\
\noalign{\smallskip}  
Adopted quantities:             &     & & \\
$P$~(days)                      & 0.798868143 & 0.798868143 & 0.798868143\\
$T$~(HJD$-$2\,400\,000)\footnote{Time of central primary eclipse} & 50770.69601 & 50770.69601 & 50770.69601\\
$e$                             &  0.00  & 0.00 & 0.00        \\ 
\noalign{\smallskip}  
Derived quantities:             &      \\
$M_p \sin^3 i$~(M$_{\sun}$)     & $1.155 \pm 0.007$ & $1.157 \pm 0.005$ & $1.157 \pm 0.006$ \\
$M_s \sin^3 i$~(M$_{\sun}$)     & $1.230 \pm 0.006$ & $1.234 \pm 0.006$ & $1.233 \pm 0.006$ \\
$a \sin i$~(R$_{\sun}$)         & $4.839 \pm 0.009$ & $4.843 \pm 0.007$ & $4.842 \pm 0.007$ \\
\noalign{\smallskip}  
Other quantities pertaining to the fit:  &      \\
$N_{obs}$                      &    18 &   18 &   18 \\ 
Time span (days)               &   793 &  793 &  793 \\ 
$\sigma_p$~(\kms)              &  1.15 & 1.26 & 1.17 \\ 
$\sigma_s$~(\kms)              &  1.42 & 0.93 & 1.14 \\ 
\noalign{\smallskip}  
\hline
\end{tabular}            
\end{minipage}
\end{table*}

\begin{figure}
\epsfxsize=095mm
\epsfbox{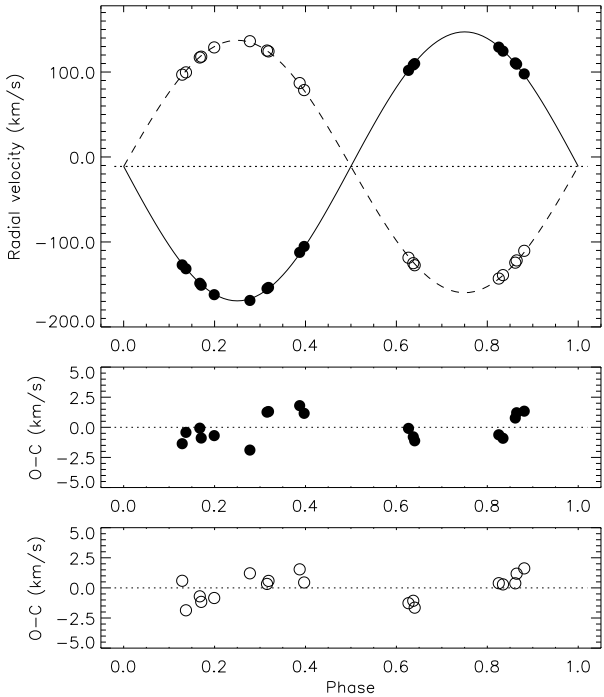}
\caption[]{\label{fig:24133_sporb}
Spectroscopic orbital solution for \24\ (solid line: primary;
dashed line: secondary) and corrected radial velocities (filled circles:
primary; open circles: secondary).
The dotted line (upper panel) represents the mean center-of-mass
velocity of the system.
Phase 0.0 corresponds to central primary eclipse.
}
\end{figure}

Spectroscopic orbits have been derived through analyses of the
radial velocities obtained from each of the two BF analyses 
of the eight selected orders.
Since the components of \24\ are quite close and deformed, the
observed light center velocities deviate somewhat from the center of
mass velocities, which are used to determine the Keplerian orbital
parameters. Before analysing the velocities, we have
therefore for each order applied phase dependent corrections as calculated 
from the Wilson-Devinney code near the corresponding wavelength range; 
see Sect.~\ref{sec:phel}. At the observed phases they
range between about $-1.2$ and +1.4 \kms\ for the primary component
and between about $-2.2$ and +1.3 \kms\ for the secondary component.
Order to order differences are less than 10\% of the corrections; 
using average corrections leads in fact to identical orbital solutions.

Next, for each observed phase, mean values of the corrected radial 
velocities from the eight selected orders were formed, and
spectroscopic orbits were then calculated using the method of
Lehman-Filh\'es implemented in the
{\sc sbop}\footnote{Spectroscopic Binary Orbit Program, \\
{\scriptsize\tt http://mintaka.sdsu.edu/faculty/etzel/}}
program (Etzel \cite{sbop}), which is a modified and expanded version of
an earlier code by Wolfe, Horak \& Storer (\cite{wolfe67}).
A circular orbit was assumed, and the period $P$ and epoch $T$ were fixed 
at the ephemeris values (Eq.~\ref{eq:24133_eph}).
Equal weights were assigned to the radial velocities,
and the two components were analysed independently (SB1 solutions).
The elements are listed in the first two columns of Table~\ref{tab:24133_spel}, 
and as seen, the results from the WD based and the simpler symmetrical BF's agree
very well, giving minimum masses accurate to about 0.6\%. 
For both set of velocities, SB2 analyses yield identical semiamplitudes.
Within errors, the system velocities agree, even without accounting for the
small difference in gravitational redshift for the components,
about 0.06 \kms.

As a further check, we have analysed the eight orders independently.
The individual semiamplitudes differ slightly more than their typical
mean errors of 0.6 \kms, but for both BF methods, their mean values agree 
very well with the results presented in Table~\ref{tab:24133_spel}.
Finally, applying instead mean radial velocities weighted according to the 
quality of the individual order solutions, 
and/or weighting the mean radial velocities according to the S/N ratio of the
observed spectra (Table~\ref{tab:feros}), lead to practically identical
elements. Also, shifting first the velocities from each order by the
difference between its system velocity and the mean system velocity
(primary and secondary components treated individually) does not change the
elements significantly.

Based on the results mentioned above, we believe that the radial velocity
differences from the two BF approaches, which for the mean values are 
within $\pm 1$ \kms, 
are more likely due to imperfections in the observed BF's, affecting the 
theoretical fits differently, than to measurable (line) 
asymmetries. Furthermore, the quality of the two datasets are comparable
with about the same order-to-order spread of the velocities. We
have therefore taken the pragmatic decision to base the final orbital elements
on the mean values of their velocities; see Table~\ref{tab:24133_rv_mean}. 
These elements are listed in the third column of Table~\ref{tab:24133_spel} 
and the corresponding orbits are shown in Fig.~\ref{fig:24133_sporb}. 
Finally, we note that if the light center velocities are applied without corrections, 
both semiamplitudes become 1.1 \kms\ smaller than listed
in Table~\ref{tab:24133_spel}, and the derived masses become about
0.03 \Msun\ lower.

Our results differ slightly from those by Rucinski et al. (\cite{r03}),
$K_p = 160.11 \pm 0.74$ \kms, $K_s = 147.21 \pm 0.63$ \kms, and
$\gamma = -12.74 \pm 0.46$ \kms, and are more accurate.

\section{Chemical abundances}
\label{sec:abund}

\begin{figure}
 \epsfxsize=095mm
\epsfbox{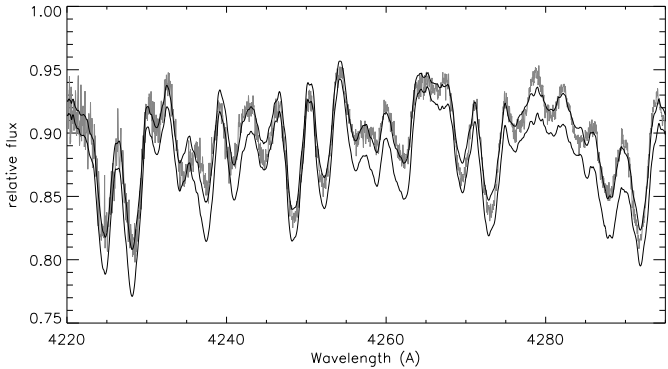}
\caption[]{\label{fig:24133_spec}
Part of FEROS spectrum taken at phase 0.168 (gray) and synthetic
binary spectra (thin) calculated for \feh\,$=-0.25$ (upper) and
\feh\,$=0.00$ (lower).
}
\end{figure}

Due to the high rotational velocities of the components 
(Sect.~\ref{sec:rv}), a detailed chemical analysis of \24\ based on the FEROS
spectra is difficult, see Fig.~\ref{fig:24133_spec}}. 
First, the lines with intrinsic equivalent widths below about 100~m\AA, 
which should preferably be used, are shallow and broad and therefore 
impossible to measure accurately. Next, line blending becomes a serious issue, 
and finally proper normalization of the spectra is difficult, especially in 
the blue spectral region.
Line by line analyses of either the observed spectra
or the reconstructed component spectra calculated from disentangled spectra
have therefore not been attempted. 
We refer to CTB08 and Clausen et al. (\cite{cbc09})
for details on line by line analyses of binaries.
We have instead established upper and lower limits for the metal abundance 
of \24\ by comparing the 
observed spectra and synthetic binary spectra calculated for a range of
scaled solar compositions. The synthetic spectra were produced as described in 
Sect.~\ref{sec:rv}. 
The overall result, based on inspection of several spectra and orders, 
is that synthetic spectra for metal abundances between -0.35 and -0.15 dex 
fit the observed spectra equally well, whereas e.g. the lines/lineblends for 
solar abundance spectra, as illustrated in Fig.~\ref{fig:24133_spec}, 
are clearly too strong.

In addition, abundances have been derived from various $uvby$ calibrations
and the indices listed in Tables~\ref{tab:24133_std} and \ref{tab:24133_absdim}.
The Holmberg et al. (\cite{holmberg07}) calibration 
gives \feh\,$=-0.25\pm0.12$ for both components, whereas
the 'blue' calibration by Nordstr\"om et al. (\cite{gencph04})
gives \feh\,$=-0.34\pm0.14$. For comparison,
the older calibration by Edvardsson et al. (\cite{be93}) gives
\feh\,$=-0.27\pm0.11$.   

In conclusion, we confirm that \24\ is (slightly) metal-weak, 
see Sect.~\ref{sec:24133}, and adopt \feh\,$=-0.25\pm0.10$.

\section{Absolute dimensions}
\label{sec:absdim}

\begin{table}   
\caption[]{\label{tab:24133_absdim}
Astrophysical data for \24.
}
\begin{minipage}{\columnwidth}
\begin{center}    
\begin{tabular}{lrr} \hline    
\noalign{\smallskip}    
\hline    
\noalign{\smallskip}    
                     &    Primary       &    Secondary      \\ 
\noalign{\smallskip}    
\hline    
\noalign{\smallskip}    
Absolute dimensions:            &                   &                 
 \\ 
$M/M_{\sun}$                    &$1.306 \pm 0.008$  &$1.392 \pm 0.008$
\\ 
$R/R_{\sun}$                    &$1.489 \pm 0.010$  &$1.782 \pm 0.011$ 
\\ 
$\log g$ (cgs)                  & $4.208 \pm 0.006$ & $4.080 \pm 0.006$
\\
$v \sin i$$^{\mathrm{a}}$ (\kms)& $92.4 \pm 1.1$    & $104.7 \pm 2.7$      
\\ 
$v_{sync}$$^{\mathrm{b}}$ (\kms)& $90.6 \pm 0.6$    & $108.5 \pm 0.7$ 
\\
 & & \\ 
Photometric data:&                       &                         \\ 
$V$         &         $7.526 \pm 0.014$  &        $ 7.128 \pm 0.011$\\  
$(b-y)$     &         $0.260 \pm 0.004$  &        $ 0.265 \pm 0.004$\\
$m_1$       &         $0.141 \pm 0.008$  &        $ 0.140 \pm 0.008$\\
$c_1$       &         $0.481 \pm 0.010$  &        $ 0.476 \pm 0.010$\\
$E(b-y)$    & \multicolumn{2}{c}   {$0.000 \pm 0.008$}  \\
$T_{\mbox{\scriptsize eff}}\,$      &  $6650 \pm  70$    &   $6625 \pm  70$ \\
$M_{\mbox{\scriptsize bol}}\,$      &  $3.27  \pm 0.05$  &   $2.89  \pm 0.05$ \\
$\log L/L_{\sun}$ & $0.59 \pm 0.02$ &    $ 0.74 \pm 0.02$ \\
$BC$              & $ 0.02$         &    $ 0.01$ \\
$M_V$ &             $ 3.25 \pm 0.05$&   $ 2.88 \pm 0.05$ \\
$V_0-M_V$          &$ 4.28  \pm 0.06 $& $ 4.25  \pm 0.06 $ \\   
Distance \, (pc) &$  71.6 \pm  2.1 $& $  70.8 \pm  2.1 $ \\
           &              &               \\
Abundance: &              &               \\
\feh\                          & \multicolumn{2}{c}   {$-0.25 \pm 0.10$} \\
\noalign{\smallskip}            
\hline
\noalign{\smallskip}
\end{tabular}            
\begin{list}{}{}
\item[$^{\mathrm{a}}$] Observed rotational velocity
\item[$^{\mathrm{b}}$] $Projected$ equatorial velocity for synchronous rotation
\end{list}
\end{center}    
\textsc{Note:}
Bolometric corrections ($BC$) by Flower (\cite{flower96}) have been
assumed, together with
$T_{eff\sun} = 5780$ K, $BC_{\sun} = -0.08$, and $M_{bol\sun} = 4.74$.
\end{minipage}
\end{table}

Absolute dimensions for the components of \24\ are calculated
from the elements given in Tables~\ref{tab:24133_phel} and
\ref{tab:24133_spel}. As seen in Table~\ref{tab:24133_absdim},
both masses and (volume) radii have been established to an 
accuracy of 0.6--0.7\%.

Individual standard $uvby$ indices are included in Table~\ref{tab:24133_absdim},
as calculated from the combined indices of \24\ outside eclipses 
(Table~\ref{tab:24133_std})
and the luminosity ratios (Table~\ref{tab:24133_phel}).
According to the calibration by Olsen (\cite{olsen88}) and the 
combined $uvby\beta$ indices at phase 0.25, there is no
significant interstellar reddening.

The adopted effective temperatures (6650~K, 6625~K) were calculated from the
calibration by Holmberg et al. (\cite{holmberg07}), assuming \feh\,$=-0.25$
(Sect.~\ref{sec:abund}). The uncertainties include those of the $uvby$ indices,
$E(b-y)$, \feh\, and the calibration itself.
Identical temperatures are obtained from the calibration by
Ram\'irez \& Mel\'endez (\cite{rm05}), whereas that by
Alonso et al. (\cite{alonso96}) leads to 100 K lower values.
2MASS photometry at phase 0.79, where $V = 6.555$, and the $V-K_s$ 
calibration by Masana et al. (\cite{masana06}) gives an average temperature
of 6600 K.

The measured rotational velocities ($v \sin i$) are close to the
projected synchronous velocities. We note that for an 
orbital inclination of 'only' $73\fdg8$, the true equatorial velocities
are about 4\% higher. 
The turbulent dissipation and radiative damping formalism of Zahn
(\cite{zahn77,zahn89}) predicts
synchronization times scales of $8.7 \times 10^5$ yr (primary) and
$3.1 \times 10^5$ yr0.1 Gyr (secondary),
and a time scale for circularization of $1.2 \times 10^7$ yr.

The distance to \24\ was calculated from the 'classical' relation
(see e.g. CTB08), adopting the solar values and bolometric corrections
given in Table~\ref{tab:24133_absdim} and accounting for all error sources.
Other $BC$ scales 
(e.g. Code et al. \cite{code76}, Bessell et al. \cite{bessell98},
Girardi et al. \cite{girardi02}) give nearly identical results.      
As seen, the distances obtained for the two components agree well.
The mean distance, 71.2 pc, which has been established to about 
3\%, is close to the result from the
new Hipparcos reduction by van Leeuwen (\cite{vl07}), $69.8 \pm 2.3$ pc, but
is marginally larger than the original Hipparcos result $65.2 \pm 3.3$ pc
(ESA \cite{hip97}).
Finally, we note that \24\ belongs to the group of eclipsing binaries within 
125 pc, discussed by Popper (\cite{dmp98}), which could be useful for 
improving the radiative flux scale.

\section{Discussion}
\label{sec:dis}

Below, we first compare the absolute dimensions obtained
for \24\ with properties of recent theoretical stellar evolutionary models, and
we then discuss \24\ together with the few other similar well-studied eclipsing 
binaries available.

\subsection{Comparison with stellar models}
\label{sec:models}

\begin{figure}
\epsfxsize=90mm
\epsfbox{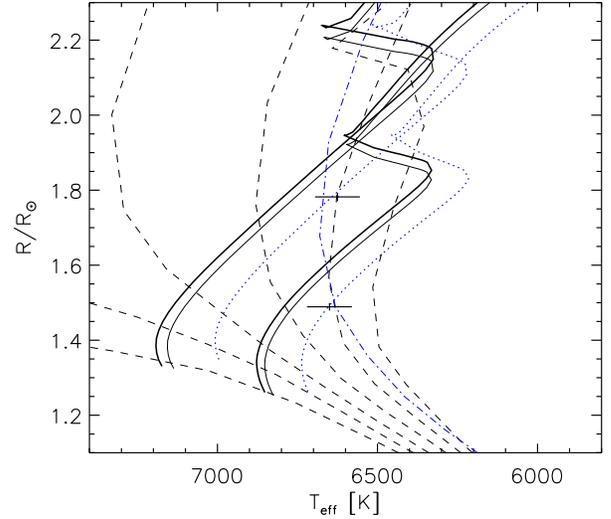}
\caption[]{\label{fig:24133_tr}
\24\ compared to $Y^2$ models
calculated for \feh\,$=-0.25$.
Tracks for the component masses (full drawn, thick) and isochrones 
for 0.5--3.0 Gyr (dashed, step 0.5 Gyr) are shown.
The uncertainty in the location of the tracks coming from
the mass errors are indicated (full drawn, thin).
To illustrate the effect of the abundance uncertainty, tracks
(dotted, blue) and the 2.2 Gyr isochrone (dash-dot, blue) 
for \feh\,$=-0.15$ are included.
}
\end{figure}

\begin{figure}
\epsfxsize=90mm
\epsfbox{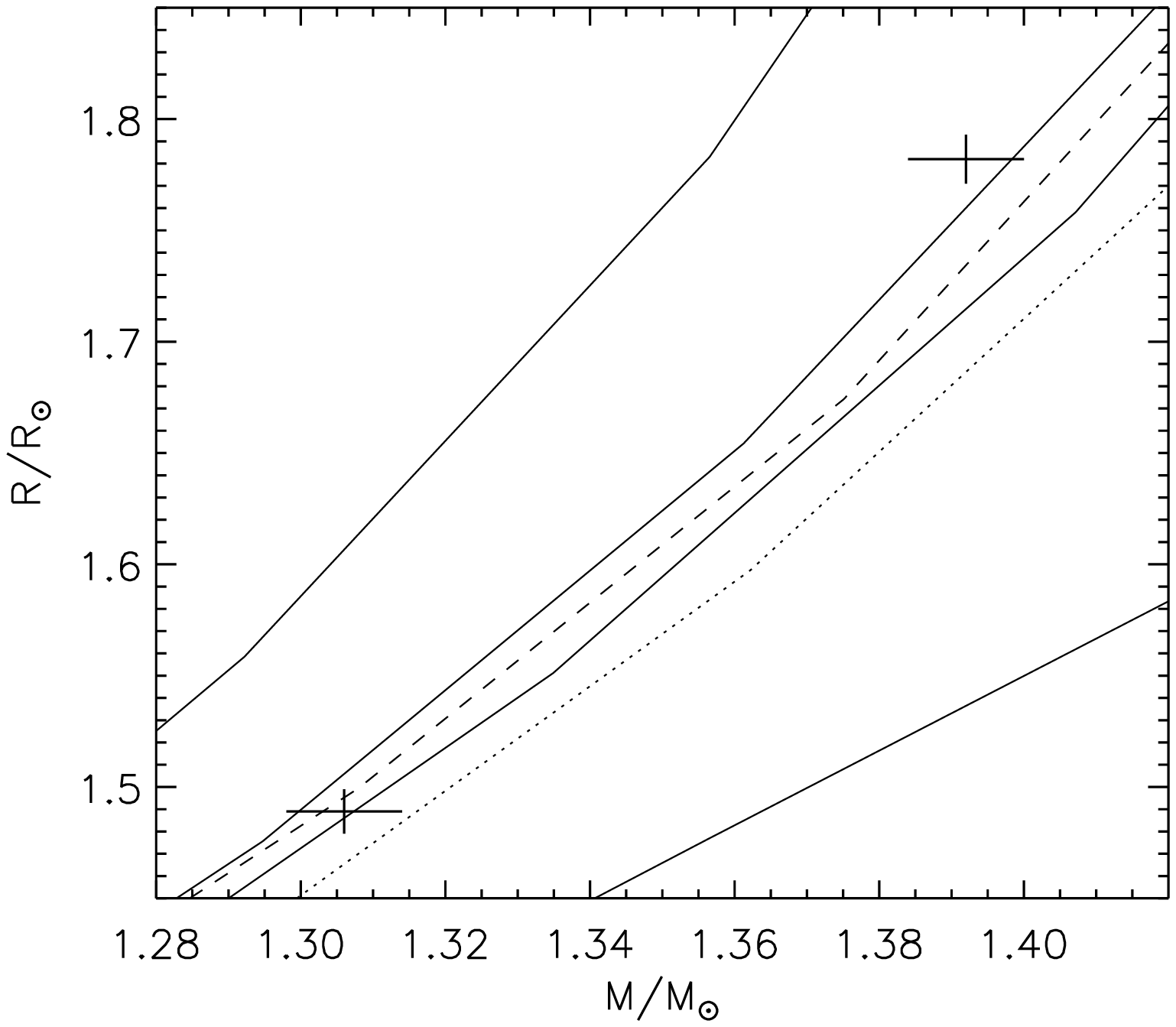}
\caption[]{\label{fig:24133_mr}
\24\ compared to $Y^2$ models
calculated for \feh\,$=-0.25$.
Isochrones (full drawn) for 1.5, 2.0, 2.13, and 2.5 Gyr are shown.
To illustrate the effect of the abundance uncertainty, 
2.0 Gyr isochrones for \feh\,$=-0.15$ (dotted) and
\feh\,$=-0.35$ (dashed) are included.
}
\end{figure}

\begin{figure}
\epsfxsize=90mm
\epsfbox{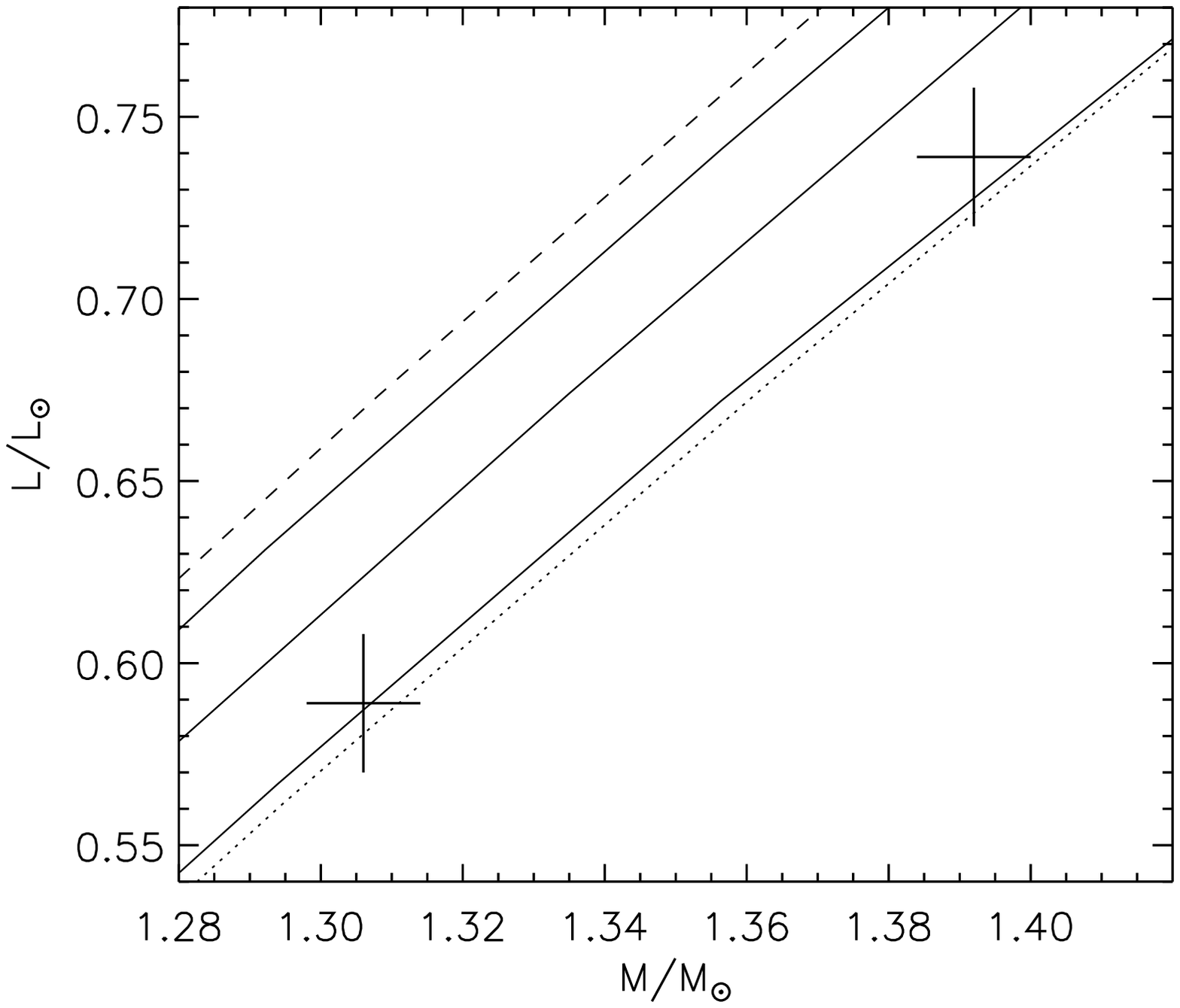}
\caption[]{\label{fig:24133_ml}
\24\ compared to $Y^2$ models
calculated for \feh\,$=-0.25$.
Isochrones (full drawn) for 1.5, 2.0, and 2.5 Gyr are shown.
To illustrate the effect of the abundance uncertainty, 
2.0 Gyr isochrones for \feh\,$=-0.15$ (dotted) and
\feh\,$=-0.35$ (dashed) are included.
}
\end{figure}

Figs.~\ref{fig:24133_tr}, \ref{fig:24133_mr},
and \ref{fig:24133_ml} illustrate the results from comparisons with the
Yonsei-Yale ($Y^2$) evolutionary tracks and isochrones by Demarque et al.
(\cite{yale04})\footnote{{\scriptsize\tt http://www.astro.yale.edu/demarque/yystar.html}}.
They include core overshoot where $\Lambda_{OS} = \lambda_{ov}/H_p$
depends on mass and also takes into account the composition dependence of 
$M_{crit}^{conv}$
\footnote{Defined as "the mass above which stars continue to have a
substantial convective core even after the end of the pre-MS phase."}.
The mixing length parameter in convective envelopes is calibrated using
the Sun, and is held fixed at $l/H_p = 1.7432$.
The enrichment law $Y = 0.23 + 2Z$ is adopted, together with the solar
mixture by Grevesse et al. (\cite{gns96}), leading to
($X$,$Y$,$Z$)$_{\sun}$ = (0.71564,0.26624,0.01812).
A brief description of other aspects of their
up-to-date input physics in given by CTB08. 
Only models for \afe\,$=0.0$ have been included in the figures.
We have used the abundance, mass, and age interpolation routines
provided by the $Y^2$ group.

\begin{figure}
\epsfxsize=90mm
\epsfbox{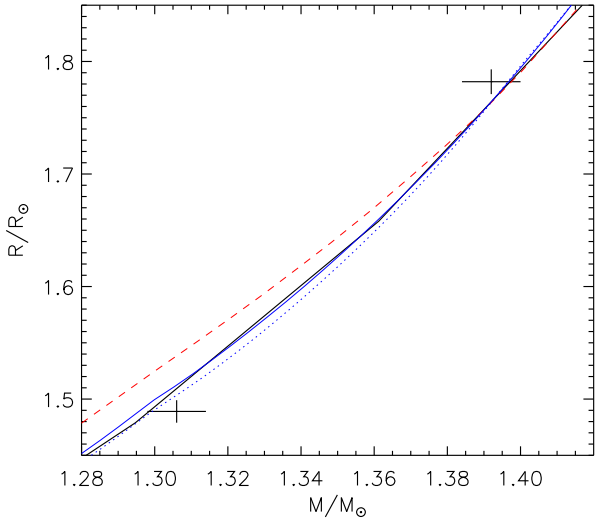}
\caption[]{\label{fig:24133_mr_mixed_av}
\24\ compared to the models listed in Table~\ref{tab:24133_models}.
Isochrones for the average ages inferred from masses and radii are shown.
$Y^2$: thick full drawn (black).
BaSTI overshoot: thin full drawn (blue).
BaSTI standard: dotted (blue).
Victoria-Regina: dashed (red).
}
\end{figure}

As seen from Fig.~\ref{fig:24133_tr}, models for the observed
masses and abundance, \feh\,$=-0.25$, equivalent to
($X$,$Y$,$Z$)$_{\sun}$ = (0.7385,0.2510,0.0105),
are about 200 K hotter than observed. The uncertainty of \feh\ is
$\pm 0.10$ dex, and tracks for \feh\,$=-0.15$, equivalent to
($X$,$Y$,$Z$) = (0.7310,0.2560,0.0130), fit the components 
at an age of about 2.2 Gyr. This can also be reached for
\feh\,$=-0.25$, if a slight hypothetical $\alpha$-element enrichment of
\afe\,$=0.15$ is introduced.
The more massive secondary component has evolved to the middle
of the main sequence band.

From a binary perspective, the most fundamental comparison
is that based on the scale-independent masses and radii, as 
shown in Fig.~\ref{fig:24133_mr}. 
The \feh\,$=-0.25$ model isochrone for 2.13 Gyr marginally fits
both components, but within the abundance uncertainty, the
general trend is that the $Y^2$ isochrones predict a higher age for 
the secondary component than for the primary. 
Although less evident, this is also seen in the mass--luminosity
diagram (Fig.~\ref{fig:24133_ml}).

\begin{table}
\caption[]{\label{tab:24133_models}
Models information and average ages inferred from masses and radii; 
see Fig.~\ref{fig:24133_mr_mixed_av}.
}
\centering
\begin{tabular}{lccccc} \hline
\hline\noalign{\smallskip}
Grid        & \feh\  &  $Y$  &  $Z$     & Age (Gyr)\\
\noalign{\smallskip}
\hline
\noalign{\smallskip}
Yonsai-Yale ($Y^2$)&$-0.25$ & 0.2510& 0.0105    & 2.13\\
Victoria-Regina    &$-0.29$ & 0.2574& 0.0100    & 1.98\\
BaSTI (overshoot)  &$-0.25$ & 0.2590& 0.0100    & 1.84\\ 
BaSTI (standard)   &$-0.25$ & 0.2590& 0.0100    & 1.80\\ 
\hline
\end{tabular}
\end{table}

In Fig.~\ref{fig:24133_mr_mixed_av} 
we have included mass-radius comparisons with the Victoria-Regina 
(VRSS grid; VandenBerg et al.,
\cite{vr06})\footnote{{\scriptsize\tt http://www1.cadc-ccda.hia-iha.nrc-cnrc.gc.ca/cvo/
community/VictoriaReginaModels/}}
and BaSTI (Pietrinferni et al.,
\cite{basti04})\footnote{{\scriptsize\tt http://www.te.astro.it/BASTI/index.php}}
models, which differ from $Y^2$, e.g. with respect to input physics,
He enrichment law, and core overshoot treatment. 
We refer to CTB08 for a brief description. Basic parameters for the models,
all with solar scaled abundances, are given in Table~\ref{tab:24133_models}.
Like the $Y^2$ models, both the standard and overshoot BaSTI models marginally
fit both components, but at a lower age.
However, the Victoria-Regina models do not fit \24\ well.
To us, this is surprising, because these models are carefully calibrated by 
cluster and binary observations. Models with \afe\,$=0.3$ (VR2A grid) can reproduce 
\24 at an age of about 2.15 Gyr, but only for \feh\ around -0.40 dex. 

\begin{table*}
\caption[]{\label{tab:24133_ac}
Information on the Claret models and ages inferred from radii;
see Figs.~\ref{fig:24133_tr_ac} and \ref{fig:24133_ar_ac}.
}
\centering
\begin{tabular}{lccccccc} \hline
\hline\noalign{\smallskip}
Model/      &  $Y$  &  $Z$  &  $l/H_p$ & $\alpha_{ov}$ & Rotation & Age (Gyr) & Age (Gyr)\\
Linestyle   &       &       &          &               &          & Primary   & Secondary\\
\noalign{\smallskip}
\hline
\noalign{\smallskip}
1 thin, blue      & 0.260 & 0.010&  1.68 & 0.20 & NO & $2.08 \pm 0.07$ & $2.24 \pm 0.03$\\
2                 & 0.260 & 0.010&  1.68 & 0.00 & NO & $2.01 \pm 0.07$ & $2.14 \pm 0.03$\\
3 dotted, blue    & 0.260 & 0.010&  1.68 & 0.20 & YES& $2.01 \pm 0.07$ & $2.19 \pm 0.03$ \\
4 dashed, blue    & 0.260 & 0.010&  1.50 & 0.20 & NO & $1.98 \pm 0.07$ & $2.20 \pm 0.03$\\
5 thick, black    & 0.240 & 0.010&  1.68 & 0.20 & NO & $2.51 \pm 0.07$ & $2.60 \pm 0.04$ \\
6                 & 0.230 & 0.010&  1.68 & 0.20 & NO & $2.75 \pm 0.07$ & $2.81 \pm 0.04$ \\
\hline
\end{tabular}
\end{table*}

Thus, except for the Victoria-Regina models, all the models with solar
scaled abundances we have tested are marginally able to reproduce \24, but
we see two general trends: First, models for the observed \feh\ are about
200 K too hot. Second, they systematically predict higher ages for 
the more massive secondary component than for the primary.
In order to look in more detail into this, we have calculated dedicated models 
for the component masses with various parameters tuned. 
For all models, we have adopted $Z$ = 0.010, which is equivalent to the 
observed \feh.  We have applied the Granada code by
Claret (\cite{c04}), which assumes an enrichment law of $Y = 0.24 + 2.0Z$
together with the solar mixture by Grevesse \& Sauval (\cite{gs98}),
leading to ($X$,$Y$,$Z$)$_{\sun}$ = (0.704,0.279,0.017).
The envelope mixing length parameter needed to reproduce the Sun 
is $l/H_p = 1.68$. The amount of core overshooting is given, in units of
the pressure scale height, by $\alpha_{ov}$.

\begin{figure}
\epsfxsize=90mm
\epsfbox{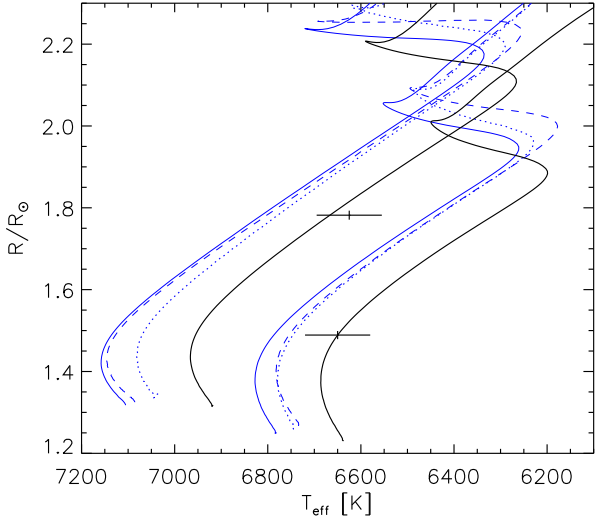}
\caption[]{\label{fig:24133_tr_ac}
\24\ compared to Claret models for the observed
masses and \feh\ abundance. See Table~\ref{tab:24133_ac}
for details and linestyles/colours.
}
\end{figure}

\begin{figure}
\epsfxsize=90mm
\epsfbox{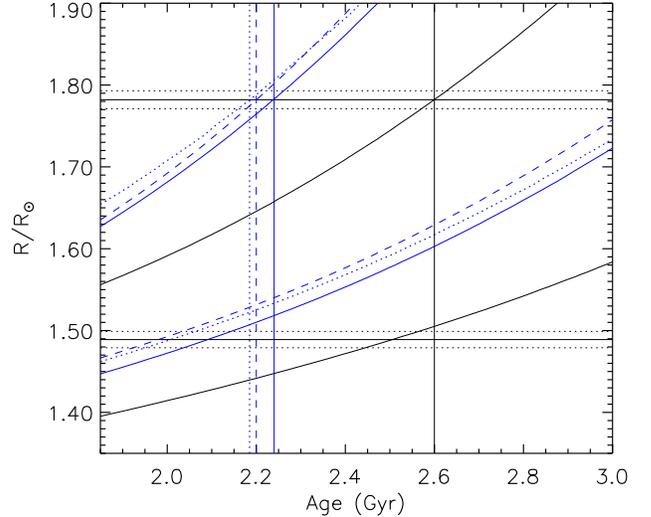}
\caption[]{\label{fig:24133_ar_ac}
\24\ compared to Claret models for the observed
masses and \feh\ abundance. See Table~\ref{tab:24133_ac}
for details.
The curves illustrate model radii as function of age for the 
components (upper: secondary; lower: primary).
The horizontal full drawn lines mark the observed radii of the
components with errors (dotted lines).
The vertical lines mark the ages predicted for the secondary
component.
}
\end{figure}

Table~\ref{tab:24133_ac} lists the models we have investigated.
As seen in Fig.~\ref{fig:24133_tr_ac}, the overshoot models (1)
with $Y$ calculated from the adopted enrichment law are too hot, as seen
for the other grids. The same is true for models without overshoot (2, not shown).
Since the components of \24\ are rotating quite
fast (Table~\ref{tab:24133_absdim}), we have calculated models which 
include rotation as described by Claret (\cite{c99}). 
Angular velocities for the models were tuned to reproduce the
observed equatorial rotational velocities of the components.
As expected, such models (3) are cooler than similar ones without rotation (1), 
but the effect is small compared to the about 200 K discrepancy.
Next, the components of \24\ have (thin) outer convection zones, and we have
therefore investigated the effect of modifying the envelope mixing length parameter. 
2D radiation hydrodynamic calculations by Ludwig et al. (\cite{hgl99}; 
see also Clausen et al. \cite{cbc09}) predict parameters, which are about
0.2 lower than for the Sun, and we have therefore adopted 1.50.
The models (4) become cooler, but again the effect is too small.
Finally, we have calculated models with a He abundance 
slightly lower than $Y = 0.26$, as given by the enrichment law for $Z = 0.01$. 
Tracks for $Y$ = 0.24 (5) and 0.23 (6, not shown) actually fit 
\24\ well. 
If we now turn to the ages, as determined from the radii, 
Fig.~\ref{fig:24133_ar_ac} shows that these models with lower $Y$ 
also predict practically identical ages for the components.
In fact, this also holds if lower $\alpha_{ov}$ values are adopted;
models without overshoot place the primary component just at the
end of the core hydrogen burning phase.  
On the other hand, all the Granada models for $Y = 0.26$ predict higher 
ages for the secondary component than for the primary, 
as seen for the other grids.

Before finishing these comparisons and drawing any definite
conclusions about
the need to adjust basic physical or chemical ingredients 
of the models, it is worth remembering that besides being fast rotating,
the components of \24\ are influenced by their mutual gravitational
and radiative interactions. They cause not only additional deformation, 
but also expansion and some heating, and 
these effects are probably somewhat different for the two stars. 
We will not elaborate further on the possible implications for the
model comparisons until additional, similar, but more detached binaries 
have been studied.

\subsection{Comparison with other binaries}
\label{sec:binaries}

Binaries like \24\, with component(s) that have evolved to the upper
half of the main sequence band, or beyond, may give important information
on core overshoot.
Already 20 years ago, such systems were found to provide 
strong evidence for convective core overshoot in intermediate mass
(1.5--2.5 \Msun) stars (Andersen et al. \cite{anc90}).
From a sample of 2--12 \Msun\ systems, Ribas et al. (\cite{rjg00})
found a significantly increasing of the amount of overshoot with 
increasing stellar mass, 
whereas Claret (\cite{c07}) found that it is less pronounced and
more uncertain. 

From the onset of core convection up to about 1.5 \Msun\ 
there are, however, only a few relevant, well-studied binaries:
(excluding active systems and systems with nearly identical components):
\object{AI\,Phe}, 
\object{BK\,Peg}, 
\object{BW\,Aqr}, and 
\object{GX Gem}. 
Andersen et al. (\cite{a88}) found that models without core overshoot
were able to reproduce AI\,Phe (1.24 + 1.20 \Msun, components above the 
main sequence) remarkably well for a normal helium
abundance\footnote{see also Torres et al. \cite{tag09}},
whereas Clausen (\cite{c91}) found that models including moderate 
overshoot gave better fits for especially the primary components of 
the slightly more massive systems BW\,Aqr (1.49 + 1.39 \Msun) and BK\,Peg
(1.43 + 1.28 \Msun). 
The latter is consistent with a lower limit of $\alpha_{ov}$ of about 0.18 for 
GX\,Gem (1.49 + 1.47 \Msun), as established by Lacy et al. (\cite{ltc08}).

We had hoped and expected, that \24\ could fill the mass gap between
these systems, but as mentioned in Sect.~\ref{sec:models} this is not the
case - Claret models with $\alpha_{ov}$ from 0 to at least 0.2 
can reproduce it perfectly well for $Y$ = 0.23--0.24.
In contradiction to this, Tomasella et al. (\cite{t08a,t08b}) report 
determination of $\alpha_{ov}$ from the much younger systems \object{V505\,Per}
and \object{V570\,Per}.

It is, however, still important to try to calibrate core overshoot
better from its onset to say 2 \Msun.
For the Victoria-Regina model grids,
VandenBerg et al. (\cite{vr06}) adopt, from binary and cluster information, 
a mass and abundance dependent amount, setting in around 1.1 \Msun\ and 
gradually increasing up to about 1.7 \Msun.
Demarque et al. (\cite{yale04}) apply a different ramping algorithm for the $Y^2$
isochrones, as do Pietrinferni et al. (\cite{basti04}) for the BaSTI 
calculations. These recipes, and others, need further empirical tests, 
and we plan to address that issue in forthcoming re-analyses of BW\,Aqr 
and BK\,Peg, which will include abundance determinations, as well as 
through new complete analyses of \object{AL\,Leo}, \object{HD76196}, 
and possibly also the \object{NGC752} member \object{DS\,And}. 

Another important aspect is the He abundance and the helium-to-metal enrichment ratio, 
and, through extrapolation, the primordial He/H abundance ratio. 
As discussed in Sect.~\ref{sec:models}, \24\
points towards a lower He abundance and/or enrichment ratio than 
the four different $Y,Z$ prescriptions adopted by the model grids studied.
We refer to Casagrande et al. (\cite{cas07}) for at recent 
determination of $\Delta Y/\Delta Z$ based on K dwarfs ($2.1 \pm 0.9$), 
to Blaser (\cite{b06}) for a HII based study ($1.41 \pm 0.62$) with references 
to a variety of methods and results, and to Ribas et al. (\cite{rjtg00})
and Claret \& Willems (\cite{cw02}) 
for determinations based on samples of eclipsing binaries ($2.2 \pm 0.8$ and
$1.9 \pm 0.6$, respectively).
We believe binaries can give an even better constraint, provided detailed
heavy element abundance determinations become available for a sufficiently large 
sample\footnote{see Torres et al. (\cite{tag09}) Table 3 
for the limited material available today}.
Such investigations are in progress for several systems, and we will
return to this matter in forthcoming papers.

Here, we close the issue with a brief historical remark: 
The use of binaries to determine the hydrogen content of stars was
pioneered by Eddington (\cite{e32}) and Str\"omgren (\cite{bs32}, \cite{bs33}),
and a few years later Str\"omgren (\cite{bs38}) also used binaries in his classical 
discussion of the helium content of the interior of the stars. 
Later, binary based helium-hydrogen abundance ratio 
determinations (for Population I stars) were published by Str\"omgren (\cite{bs67}) 
and Popper et al. (\cite{dmp70}).

\section{Summary and conclusions}
\label{sec:sum}
From state-of-the-art observations and analyses, precise (0.6--0.7\%)
absolute dimensions have been established for the nearby, early F-type, double-lined,
detached eclipsing binary \24. From synthetic spectra and $uvby$ calibrations,
a metal abundance of \feh\,$=-0.25\pm0.10$ has been derived.
The 1.39 \Msun\ secondary component has evolved to the middle of the main-sequence
band and is slightly cooler than the 1.31 \Msun\ primary.
The $P = 0\fd80$ period orbit is circular and the observed rotational velocities
of the components, $92.4 \pm 1.1$ (primary) and $104.7 \pm 2.7$ (secondary) 
\kms, correspond closely to synchronization.

Yonsai-Yale, BaSTI, and Granada evolutionary models for the observed
metal abundance and a 'normal' He content of $Y = 0.25-0.26$, 
as established from the adopted helium enrichment laws, 
marginally reproduce the components at ages between 1.8 and 2.1 Gyr.
All such models are, however, systematically about 200 K hotter than observed, and 
predict ages for the more massive component, which are systematically
higher than for the less massive component. The latter is even more
pronounced for Victoria-Regina models.
The two trends can not be removed by adjusting the amount of core overshoot or envelope 
convection level, or by including rotation in the model calculations.
They may be due to proximity effects in \24, but
on the other hand, we find excellent agreement for 2.5--2.8 Gyr Granada models with a 
slightly lower $Y$ of 0.23--0.24. 

We had expected that \24\ is sufficiently evolved to provide new information on the 
level of core overshoot in the 1.1--1.7 \Msun interval, where it is believed to 
ramp up, but this is not the case.
\24\ can be reproduced by models calculated for $\alpha_{ov}$ from 0.0 to at 
least 0.2. The preference for a helium content of 0.23--0.24 is interesting,
but more well-detached systems with measured metal abundances are needed for 
any firm conclusions on the implications for example for the helium enrichment law. 
We will return to these issues in forthcoming papers on 
other systems recently observed within the Copenhagen binary project.

\begin{acknowledgements}
We thank B.~R. J{\o}rgensen, J. Mouette, and N.~T.
Kaltcheva for participating in the (semi)automatic observations 
of \24\ at the SAT.
A. Kaufer, O. Stahl, S. Tubbesing, and B. Wolf kindly obtained
seven FEROS spectra of \24\ during Heidelberg/Copenhagen guaranteed
time in 1999.
Excellent technical support was received from the staffs of Copenhagen
University and ESO, La Silla.
We thank J.~M. Kreiner for providing a complete list of published times
of eclipse for \24, H. Bruntt for making his $bssynth$ software available,
and J. Southworth for access to his JKTWD code.

The projects "Stellar structure and evolution -- new challenges from
ground and space observations" and "Stars: Central engines of the evolution
of the Universe", carried out at Copenhagen University and Aarhus University,
are supported by the Danish National Science Research Council.

The following internet-based resources were used in research for
this paper: the NASA Astrophysics Data System; the SIMBAD database
and the VizieR service operated by CDS, Strasbourg, France; the
ar$\chi$iv scientific paper preprint service operated by Cornell University;
the VALD database made available through the Institute of Astronomy,
Vienna, Austria.
This publication makes use of data products from the Two Micron
All Sky Survey, which is a joint project of the University of
Massachusetts and the Infrared Processing and Analysis Center/
California Institute of Technology, funded by the National
Aeronautics and Space Administration and the National Science
Foundation.
\end{acknowledgements}

{}

\listofobjects
\begin{appendix}
\section{Radial velocity observations}
\label{sec:rvtab}
\begin{table*}
\caption[]{\label{tab:24133_rv_mean}
Mean radial velocities for \24. 
}
\begin{center}
\begin{tabular}{lrrrrrrr}
\hline
\hline\noalign{\smallskip}
HJD $-$ 2\,400\,000& Phase    & $RV_p$$^{\mathrm{a}}$ & $\delta RV_p$$^{\mathrm{b}}$ &  $O-C$$^{\mathrm{c}}$ & $RV_s$$^{\mathrm{a}}$ & $\delta RV_s$$^{\mathrm{b}}$ &  $O-C$$^{\mathrm{c}}$  \\
\noalign{\smallskip}
\hline
\noalign{\smallskip}
51188.63785 & 0.16750 & $-147.46$ & $ 1.16$  & $-0.07$ & $ 115.92$ & $-1.16$ & $-0.70$ \\
51207.54462 & 0.83445 & $ 123.44$ & $-1.14$  & $-0.91$ & $-137.91$ & $ 1.12$ & $ 0.29$ \\
51207.56623 & 0.86150 & $ 109.88$ & $-0.86$  & $ 0.77$ & $-123.77$ & $ 0.60$ & $ 0.38$ \\
51208.58523 & 0.13705 & $-130.70$ & $ 0.84$  & $-0.42$ & $  99.00$ & $-0.58$ & $-1.86$ \\
51209.59181 & 0.39706 & $-105.16$ & $ 0.12$  & $ 1.16$ & $  79.91$ & $ 1.18$ & $ 0.45$ \\
51211.56298 & 0.86451 & $ 108.40$ & $-0.83$  & $ 1.21$ & $-121.21$ & $ 0.55$ & $ 1.18$ \\
51212.60641 & 0.17065 & $-149.78$ & $ 1.18$  & $-0.89$ & $ 116.82$ & $-1.23$ & $-1.16$ \\
51385.92739 & 0.12883 & $-126.25$ & $ 0.75$  & $-1.36$ & $  96.42$ & $-0.45$ & $ 0.59$ \\
51386.93266 & 0.38720 & $-112.11$ & $ 0.18$  & $ 1.80$ & $  87.83$ & $ 0.84$ & $ 1.54$ \\
51390.86963 & 0.31539 & $-154.09$ & $ 0.83$  & $ 1.26$ & $ 123.86$ & $-1.39$ & $ 0.33$ \\
51391.91712 & 0.62660 & $ 101.65$ & $-0.27$  & $-0.09$ & $-119.05$ & $-0.49$ & $-1.27$ \\
51392.87464 & 0.82520 & $ 128.06$ & $-1.22$  & $-0.62$ & $-141.76$ & $ 1.32$ & $ 0.39$ \\
51562.53347 & 0.19921 & $-160.66$ & $ 1.39$  & $-0.69$ & $ 127.13$ & $-1.82$ & $-0.84$ \\
51562.59618 & 0.27771 & $-167.63$ & $ 1.21$  & $-1.89$ & $ 134.03$ & $-2.23$ & $ 1.21$ \\
51562.62891 & 0.31868 & $-152.73$ & $ 0.79$  & $ 1.31$ & $ 122.97$ & $-1.29$ & $ 0.58$ \\
51977.49585 & 0.63709 & $ 107.92$ & $-0.35$  & $-0.79$ & $-125.16$ & $-0.17$ & $-1.07$ \\
51978.48974 & 0.88122 & $  97.12$ & $-0.65$  & $ 1.34$ & $-110.12$ & $ 0.31$ & $ 1.62$ \\
51981.49266 & 0.64018 & $ 109.54$ & $-0.38$  & $-1.12$ & $-127.48$ & $-0.07$ & $-1.64$ \\
\hline
\end{tabular}
\begin{list}{}{}
\item[$^{\mathrm{a}}$] Mean values of the {\it measured} radial velocities (\kms)
\item[$^{\mathrm{b}}$] Mean correstions. Approximate center of mass velocities are 
obtained by subtracting these corrections (\kms) from the measured velocities.
For the orbital solutions, individual corrections were used for each order.
\item[$^{\mathrm{c}}$] $O-C$ residuals (\kms) from the adopted spectroscopic orbits 
(Table~\ref{tab:24133_spel}).
\end{list}
\end{center}
\end{table*}
   
\end{appendix}
\end{document}